# Design and realization of Ohmic and Schottky interfaces for oxide electronics

*Jie Zhang*[1], *Yun-Yi Pai*[1], *Jason Lapano*[1], *Alessandro R. Mazza*[1], *Ho Nyung Lee*[1], *Rob Moore*[1], *Benjamin J. Lawrie*[1], *T. Zac Ward*[1], *Gyula Eres*[1], *Valentino R. Cooper*[1#], *Matthew Brahlek*[1]*

[1]Materials Science and Technology Division, Oak Ridge National Laboratory, Oak Ridge, Tennessee 37831, USA

Email: *brahlekm@ornl.gov, #coopervr@ornl.gov

**Abstract:** Understanding band alignment and charge transfer at complex oxide interfaces is critical to tailoring and utilizing their diverse functionality. Towards this goal, we design and experimentally validate both Ohmic- and Schottky-like charge transfers at oxide/oxide semiconductor/metal interfaces. We utilize a method for predicting band alignment and charge transfer in $AB$O$_3$ perovskites, where previously established rules for simple semiconductors fail. The prototypical systems chosen are the rare class of oxide metals, Sr$B$O$_3$ with $B$=V-Ta, when interfaced with the multifaceted semiconducting oxide, SrTiO$_3$. For $B$=Nb and Ta, we confirm that a large accumulation of charge occurs in SrTiO$_3$ due to higher energy Nb and Ta $d$-states relative to Ti; this gives rise to a high mobility metallic interface, which is an ideal epitaxial oxide/oxide Ohmic contact. On the other hand, for $B$=V, there is no charge transfer into the SrTiO$_3$ interface, which serves as a highly conductive epitaxial gate metal. Going beyond these specific cases, this work opens the door to integrating the vast phenomena of $AB$O$_3$ perovskites into a wide range of practical devices.

This manuscript has been authored by UT-Battelle, LLC under Contract No. DE-AC05-00OR22725 with the U.S. Department of Energy. The United States Government retains and the publisher, by accepting the article for publication, acknowledges that the United States Government retains a non-exclusive, paid-up, irrevocable, world-wide license to publish or reproduce the published form of this manuscript, or allow others to do so, for United States Government purposes. The Department of Energy will provide public access to these results of federally sponsored research in accordance with the DOE Public Access Plan (http://energy.gov/downloads/doe-public-access-plan).

The last half-century has seen nearly exponential growth in electronic-based technologies that have permeated and changed the daily lives of nearly all humans. This was driven by the combination of downscaling of the transistor with simultaneous cost reductions and the rapid evolution of a wide array of peripheral devices such as displays, batteries, data transmission and storage, etc. Currently, however, there are critical obstacles to continued innovation related to fundamental physical size limitations [1,2], as well as societal needs such as energy efficient devices and more robust energy harvesting approaches. Many promising directions to overcome these hurdles are focused on diversifying materials to include systems with functionality beyond traditional metals, insulators, and semiconductors. The $ABO_3$ perovskite oxides are viewed as key candidates to compliment or replace current materials in new electronic devices due to the large array of functional properties[3–5]. These novel aspects are directly tied to the activation of multiple electron degrees of freedom (spin, orbital and charge), and the much smaller characteristic length scales, such as short screening lengths associated with strong non-linear dielectric responses at highly charged interfaces, strong localized magnetism and Coulombic interactions. Furthermore, there are a huge number of elemental combinations that fit onto the perovskite cation sublattices, which exhibit both traditional functionalities as well as a wide array of additional physical properties; spanning conventional semiconductors, insulators, and metals, to more exotic systems including a wide variety of magnetic materials, ferroelectric, superconductors, as well as metastable phase-change materials. These have implications across areas ranging from applications in dielectrics and power electronics, to spintronics, neuromorphic computing and even superconducting based quantum technologies.

A key difficulty in utilizing the vast functionality of complex oxides is forming structurally and chemically stable heterojunctions, which enable processes such as carrier injection and depletion that form the basis of ubiquitous Ohmic and Schottky barriers. Due to the high offset of the work functions, many noble metals and metal alloys[6] form only Schottky barriers in conjunction with prototypical perovskite materials like $SrTiO_3$[7–9]. Lower work function metals are typically more reactive and adhere better to oxide surfaces, but they often form interfacial oxide layers or intermix with the oxide. For example, key materials that form Ohmic contacts, such as Cr, do so partially because of interfacial diffusion[10]. Furthermore, epitaxial integration of metals with perovskites is rare. For example, Pt, Cu (which both form Schottky junctions with $SrTiO_3$) and Cr orient the [001] // to the $SrTiO_3$ [001], while Ag and Au form textured films[10–12]. The mismatch between metals and oxides often leads to challenges related to epitaxial relaxation and defect formation, unintended chemical interactions often due to high oxygen diffusion from the perovskite under typical growth temperatures, or non-ideal effects related to texturing or high interfacial resistance.[10,13] Oxide-to-oxide type junctions provide a promising avenue to overcome these issues and give access to a more diverse palette of materials for the design of both high-quality Ohmic and Schottky barriers[10]. This approach also enables new levels of control through epitaxy, which may be applied to realize designer functionalities not accessible in simple metal-to-oxide interfaces[14]. Identifying a wide range of functionally diverse and robustly compatible oxide candidates that can be epitaxially coupled without disturbing the sharp atomic interface is a critical step towards device integration in complex oxides.[15] However, Anderson's and the Schottky-Mott rules, empirically used to predict band alignment and charge transfer based on the energetics of the bulk compounds relative to the vacuum,[16] are of limited use for applications in perovskites. This is due to invalid assumptions in general[17], and further complications that arise at oxide/oxide interfaces such as the fundamentally smaller characteristic length scales of charge transfer and microscopic complexities due to bond distortions and atomic rearrangement, and the invalidity of the assumption of a single-valued work function [9,18,19].

Towards this goal, we design and realize two key interface types for electronic devices, the Ohmic- and Schottky-type interfaces at epitaxial oxide/oxide semiconductor/metal junctions. This is achieved by utilizing a new physically accurate band alignment scheme based on the continuity of the oxygen 2*p* states[18]. This work centers on the versatile and most widely used perovskite semiconductor $SrTiO_3$, with the VB-group Sr(V-Ta)$O_3$ oxides which are one of the few highly conductive metal perovskites. We predict and experimentally validate that $SrTiO_3$/$SrVO_3$ forms a Schottky-type interface, while $SrTiO_3$/$SrTaO_3$ is Ohmic. Surprisingly, we find that the $SrTaO_3$/$SrTiO_3$ system gives rise to a conductive state with a residual resistivity ratio (*RRR*) on the order of 10,000, high mobility, and large magnetoresistance (MR), which makes it an unusually good Ohmic contact material. We demonstrate that the band alignment is dictated by the oxygen continuum boundary condition, which serves as a powerful tool for designing all-oxide devices around a wide array of novel functional phenomena. We survey an important collection of materials combinations, which will act as a guide to finding and creating new functional devices for next generation applications.

A recently developed scheme by Zhong and Hansmann for predicting the sign and amplitude of the charge transfer at oxide interfaces is directly associated with the relative energy difference of the oxygen 2*p* (O2*p*) states in the $ABO_3$/$AB'O_3$ heterojunctions [18], as illustrated in Figure 1(a) for the specific cases of $SrTiO_3$ and $SrVO_3$. To establish equilibrium, i.e. a constant Fermi energy, charge must be transferred across the interface. As schematically depicted in Figure 1(b), when the O2*p* states are aligned the direction of charge transfer is dictated by the difference in Fermi energies, $E_F$, relative to the aligned O2*p* states, which is denoted by $\epsilon_p$. Applying this theory would suggest, for the specific case here, that since $\epsilon_{p,SrTaO3}$ - $\epsilon_{p,SrTiO3}$ > 0, the charge will be transferred from the $SrTaO_3$ to the $SrTiO_3$, thereby populating the states on the $SrTiO_3$ side, as shown in Figure 1(c). Alternatively, for the case of $\epsilon_{p,SrVO3}$ - $\epsilon_{p,SrTiO3}$ < 0 charge associated with any occupied states on the $SrTiO_3$ side will be transferred to the $SrVO_3$, thereby depopulating the $SrTiO_3$.

The interfaces among the semiconductor $SrTiO_3$ with VB-group Sr(V-Ta)$O_3$ (semiconductor/metal) provides an ideal testbed to probe the physics of band alignment at oxide interfaces and is of high-relevance from an oxide-electronics perspective. Within the perovskites, good metals are rare, which highlights the importance of the VB group Sr(V-Ta)$O_3$. When going from V to Ta, the work function difference is large enough to cover the energy range needed to form both Schottky ($SrVO_3$) and Ohmic ($SrTaO_3$) heterojunctions with $SrTiO_3$, as schematically shown in Figure 1(b) and (c). Additionally, the electronic configuration in the Sr(V-Ta)$O_3$ system is relatively simple and does not involve complexities such as magnetism, as is common for the most well-known perovskite metal, $SrRuO_3$. The epitaxial synthesis of $SrVO_3$ and $SrTaO_3$ is relatively easy due to their crystalline compatibility with $SrTiO_3$, since both have simple cubic structures with close to unity tolerance factor[20,21] and high chemical stability. Lastly, Sr(V-Ta)$O_3$ perovskites represent an important class of functionality where the strong electron-electron correlations open a region of transparency in the visible to ultraviolet portions of the spectrum,[22–24] which is uncommon for metals in general.

To go beyond this schematic argument, we performed first principles calculations for the $SrTaO_3$/$SrTiO_3$ and $SrVO_3$/$SrTiO_3$ systems with the results being shown in Figure 2. The energy dependent density of states (DOS) was calculated for all three systems in a bulk-like configuration and the energy was shifted to align the weighted-band center of the O2*p* states, as shown in Figure 2(a). Here we show the Fermi energy for $SrTiO_3$ in the intrinsic limit (i.e. the Fermi level is within the gap in the limit of no charge doping), as well as near the bottom of the conduction band, which is the experimentally relevant point since bulk $SrTiO_3$ ubiquitously exhibits electron-doped oxygen vacancies. (This rigid-band shift for bulk $SrTiO_3$ is valid in the dilute limit; however, at interfaces there may be a larger density of oxygen vacancies that may locally distort the O2$p$ states, which could modify the charge transfer profile). Based on the relative position of the Fermi energies (i.e. the relative values of $\epsilon_p$) it can be surmised that SrTaO$_3$ will transfer charge to SrTiO$_3$, thus creating an accumulation layer (identical results are found for SrNbO$_3$, see Supporting Information). For SrVO$_3$, we find that the values of $\epsilon_p$ are relatively close to that of SrTiO$_3$, which necessitates a more realistic set of calculations to determine the charge transfer direction. Therefore, we performed heterostructure calculations where 7 unit cells of both SrTaO$_3$ and SrVO$_3$ were interfaced with 7 unit cells of SrTiO$_3$. The results for these calculations are shown in Figure 2(b) where the number of free electrons (related to the layer-resolved density of states) was determined for the individual atomic layers across the entirety of the heterostructure. For both systems the number of electrons on the left side is non-zero over the entirety of the thickness, which is consistent with the metallic nature of these materials. However, for the SrTiO$_3$ side, there is a significant difference. For SrTaO$_3$/SrTiO$_3$, the number of electrons on the SrTiO$_3$ side is non-zero, which indicates that there is an emergent 2-dimensional electron gas (2DEG) due to charge transfer[25–28]. Whereas, for the SrVO$_3$/SrTiO$_3$, the number of electrons on the SrTiO$_3$ side is zero, indicating there is no charge transfer into the SrTiO$_3$ and no 2DEG exists. The important feature here, is that unlike the Sr(Ta/Nb)O$_3$/SrTiO$_3$ superlattices, in the case of SrVO$_3$/SrTiO$_3$ the occupied V-$d$ states are lower in energy relative to the center of the oxygen band than the Ti-$d$ states. This would suggest that the electrons should remain in the V-$d$ bands rather than being transferred to the Ti-$d$ states. Therefore, first principles calculations predict that SrVO$_3$-SrTaO$_3$ on SrTiO$_3$ span Ohmic-like to Schottky-like charge transfer and are experimentally accessible.

Our first principles calculations have shown that charge transfers into SrTiO$_3$ from Sr(Ta/Nb)O$_3$ due to the relative band alignment, which gives rise to an interfacial 2DEG. This provides a simple route to test the band alignment of SrTiO$_3$ relative to SrVO$_3$ and SrTaO$_3$ using simple low temperature transport. Since SrVO$_3$ does not transfer charge into SrTiO$_3$, no 2DEG should form and conduction should be solely through the SrVO$_3$ film. On the other hand, in the case of SrTaO$_3$ a 2DEG should form, which will conduct in parallel with the SrTaO$_3$. 2DEGs in SrTiO$_3$ are characterized by extremely high mobilities, in excess of 100,000 cm$^2$V$^{-1}$s$^{-1}$, carrier densities in the range of 10$^{13}$-10$^{15}$ cm$^{-2}$/10$^{19}$-10$^{20}$ cm$^{-3}$[29], and an accumulation length that is typically of the scale of 5 nm dictated by the high dielectric constant of SrTiO$_3$. This combination of high mobility and high interfacial carrier density gives rise to very large conductivity, which, may be comparable to the conductivity of the perovskite metal (SrTaO$_3$). This is due to the fact that, despite the carrier density in SrTaO$_3$ being of order 10$^{22}$ cm$^{-3}$[30], mobilities are typically quite low for pulsed laser deposition (PLD) grown perovskite metal films due to a large effective mass combined with short scattering times due to typical dominance of strong surface scattering[31], uncontrolled non-stoichiometry[32], or both.

Figure 3 shows a compilation of temperature dependent transport measurements for SrVO$_3$ and SrTaO$_3$ films grown by PLD on SrTiO$_3$ as well as highly insulating LSAT and GdScO$_3$. Here the films were grown using a KrF excimer laser with 5 Hz repetition rate, a fluence of 0.5 J/cm$^2$ at a temperature of 650 °C in oxygen partial pressure of 4×10$^{-6}$ Torr (see Supporting Information section 1 for details regarding the synthesis). The structure was characterize using X-ray diffraction and found to be epitaxially strained for all films. The transport measurements were performed in the standard van der Pauw geometry with Ohmic indium contacts. Figure 3(a), shows the temperature dependence of the resistivity for both SrVO$_3$ and SrTaO$_3$ on SrTiO$_3$. This is obtained by multiplying the sheet resistance, $R_S$, multiplied by the film thickness ($R_S$ is indicated in Figure 3(a) and Supporting Information Figure S4). Both materials exhibit resistivities that decrease with decreasing temperature, which indicates that both systems are metals. Although qualitatively similar, the two systems are quantitatively different in that the low temperature resistivity of the SrTaO$_3$ system is orders-of-magnitude smaller than the SrVO$_3$ system. SrVO$_3$ has a low temperature resistivity of $2\times10^{-4}$ $\Omega$cm, whereas the SrTaO$_3$/SrTiO$_3$ system exhibits a nominal low temperature resistivity 4 orders of magnitude smaller at around $8\times10^{-9}$ $\Omega$cm. In contrast, the room temperature resistivities are similar at around $1\times10^{-4}$ $\Omega$cm. It is customary to use the residual resistivity ratio, $RRR=\rho(T=300K)/\rho(T=2K)$, as a metric of the concentration of defects within a materials class, which can be seen in Figure 3(b) where $\rho(T=300K)/\rho(T)$ is plotted. For example, hybrid molecular beam epitaxy-grown SrVO$_3$ films with a high level of control over the stoichiometry can reach a RRR over 200,[20,33,34] which is likely limited only by surface scattering. In contrast, films grown by PLD typically possess lower RRRs due to higher defect concentrations incorporated during synthesis. Our PLD grown SrVO$_3$ exhibits RRR on the order of 1 when grown on SrTiO$_3$ and ~4 when grown on highly insulating LSAT (the slightly larger RRR may be due to better lattice matching on LSAT than on SrTiO$_3$). In contrast, SrTaO$_3$ grown under identical conditions, exhibits RRRs around 10,000, as shown in Figure 3(b). It is tempting to conclude that the SrTaO$_3$ heterostructure is of much higher quality than the SrVO$_3$. However, growing identical SrTaO$_3$ films on highly insulating GdScO$_3$ reveals a RRR similar to that of SrVO$_3$. This indicates that the single-slab-interpretation is not valid. Instead, one must consider the entire interfacial SrVO$_3$/SrTiO$_3$ and SrTaO$_3$/SrTiO$_3$ systems.

To see that the interface with SrTiO$_3$ must be playing a role, we consider the resistivities of SrTaO$_3$ grown on highly-insulating GdScO$_3$ relative to that of bulk SrTiO$_3$ versus doping levels[35,36]. The resistivity of SrTaO$_3$ on GdScO$_3$ is found to be nearly temperature independent at around $1\times10^{-4}$ $\Omega$cm, several orders larger than on SrTiO$_3$. The low temperature resistivity in bulk SrTiO$_3$ (doped with either Nb or oxygen vacancies) is typically found to have a minimum at $10^{-3}$-$10^{-4}$ $\Omega$cm at a doping greater than $10^{16}$-$10^{17}$ cm$^{-3}$. Carrier densities due to oxygen vacancies of this range can easily be achieved by annealing SrTiO$_3$ in vacuum at temperatures similar to those used to grow the films in this work, 650 °C. Therefore, assuming that the band alignment theory is correct implies that the transport is likely through the SrTaO$_3$ and the SrTaO$_3$/SrTiO$_3$ interface, and possibly the bulk SrTiO$_3$. If this is so, the measured resistivity would be given by the summation of the parallel conductance channels, $G_{Total} = G_{SrTiO3} + G_{2DEG} + G_{SrTaO3}$, where each conductance channel is given by $e\mu n_{2D}$, where $e$ is the electron charge, $\mu$ the mobility and $n_{2D}$ is the areal carrier density of each channel ($n_{3D}\times thickness$), respectively. Experimentally, measuring the net resistivity is given by $G_{Total}^{-1}$ multiplied by the thickness of SrTaO$_3$. Given typical mobilities and carrier densities for bulk SrTiO$_3$ and interfacial 2DEGs, this yields effective resistivities on the order of $10^{-8}$-$10^{-9}$ $\Omega$cm at 2 K and $10^{-4}$ at room temperature, which identically matches the experimental data shown in Figure 3. Since the SrVO$_3$ and SrTaO$_3$ systems were grown under identical conditions (temperature and oxygen partial pressures), they should have identical SrTiO$_3$ doping levels. Thus, we can conclude that the SrTaO$_3$/SrTiO$_3$ system forms an Ohmic junction, whereas the SrVO$_3$/SrTiO$_3$ system forms a Schottky junction.

Magnetotransport measurements give additional insights into the origin and character of the strikingly different behaviors of the SrVO$_3$/SrTiO$_3$ and SrTaO$_3$/SrTiO$_3$ systems and point to the formation of an interfacial 2DEG. The magnetoresistance ($MR=[\rho(B)-\rho(B=0T)]/\rho(B=0T)\times100\%$) is shown in Figure 4(a) and (b) and the Hall resistance is shown in Figures 4(c) and (d) for both SrVO$_3$/SrTiO$_3$ and SrTaO$_3$/SrTiO$_3$, respectively. We first discuss the MR then the Hall effect. Due to the orbital motion, a semiclassical picture of the magnetoresistance of a metal (strictly in a multiband case) is a function of the mobility, $MR\sim(\mu B)^2$ (see Ref. [37]), which is in the low field limit ($\mu B<1$). However, for higher magnetic fields the MR should saturate for a semiclassical model or continue to grow unbounded if more complex physics is at play.[37] Here, we see that the SrVO$_3$ system exhibits only low-field orbital MR ($MR\sim(\mu B)^2$), which is maximum at 0.10% at 9 T and 2 K. This reflects the low film mobility and the absence of any substrate effects. In contrast, the behavior of SrTaO$_3$ is more complex. In the low field limit, $MR\sim(\mu B)^2$ is dominated by the orbital motion at all temperatures. For low temperatures, the MR rises to 1000% at around 1-2 T, which is consistent with mobilities on the order of 5,000-10,000 cm$^2$V$^{-1}$s$^{-1}$. In the high-field limit, however, the MR grows roughly linearly to the maximum field used, 9 T, to a value of nearly 40,000%. To fully explain this behavior requires more complex transport models and will be a question of future interest. Nevertheless, this data points to the combination of a high-mobility state in conjunction with a low mobility state, which can be further elucidated in the context of the Hall effect data.

For a semiclassical single-band metal the Hall slope is a direct measure of the density of charge carriers. In the limit of multiple bands or systems with complex Fermi surface topologies, the interpretation is more intricate but can provide a semi-quantitative characterization of the metallic state. SrVO$_3$, SrTaO$_3$, bulk SrTiO$_3$, and 2DEG systems can exhibit multiband effects only when the mobility is sufficiently high. In the SrVO$_3$ system, $R_{xy}$ is linear and temperature independent, with a slope that yields a carrier density that is close to the single band carrier density (1 electron per unit cell ~10$^{22}$ cm$^{-3}$). This reflects the overall low mobility and the fact that the transport is solely through the SrVO$_3$ film. In contrast, $R_{xy}$ for SrTaO$_3$ is non-linear and strongly temperature dependent. This is indicative of multiple carriers with disparate carrier densities and mobilities. The qualitative field dependence of $R_{xy}$ (low field slope smaller than the high-field slope) indicates one of the dominant carriers to be hole-like, which, despite no physical hole-like pockets, can be attributed to portions of the Fermi surface with negative curvature[38,39]. As discussed in the Supporting Information, the field and temperature dependence of the Hall effect can be modeled as a three-carrier system composed of low mobility SrTaO$_3$, the high mobility SrTiO$_3$ as well as an interfacial 2DEG. The hole-like band is chosen to be the 2DEG, because bulk SrTiO$_3$ in the carrier density range of 10$^{16}$-10$^{17}$ cm$^{-3}$ is only known to exhibit electron-like Hall slopes (see, for example, Ref. [40]). However, high-mobility 2DEG systems on SrTiO$_3$ commonly exhibit both electron- and hole-like multicarrier Hall slopes, which likely point to a complex interfacial Fermi surface which is electronically active and plays a non-negligible effect in the overall conductance. The results of the fitting yield an SrTiO$_3$ carrier density in the range of 10$^{16}$ cm$^{-3}$, which exactly matches the expected carrier range for the growth temperatures used[35]. Furthermore, the carrier density for the 2DEG is in the range of 10$^{13}$ cm$^{-2}$, which is again consistent with both the expected charge transfer at an oxide interface as well as with known 2DEGs on SrTiO$_3$. The mobilities are of the range of several thousand which match previous measurements.

The SrTaO$_3$/SrTiO$_3$ system exhibits a rich array of transport phenomena that indicate the conduction is through both the SrTaO$_3$, an interfacial 2DEG, and the bulk SrTiO$_3$ as would be expected in an Ohmic-type contact. The SrVO$_3$/SrTiO$_3$ system presents more straightforward transport solely through the SrVO$_3$ layer as expected for a Schottky contact. Ohmic point contacts with a diameter of <500 µm were applied to the film surface and back of the substrate to conduct current-voltage (I-V) measurements across the interface for both systems grown on Nb-doped SrTiO$_3$ substrates with doping at 0.5%. The results are shown in the insets of Figure 4(c and d). Here, the I-V curves for SrVO$_3$/SrTiO$_3$ show an asymmetric character with a much steeper slope on the positive bias side. This is a classic indicator of Schottky-like charge transfer. In contrast, the SrTaO$_3$ shows a symmetric and nearly linear I-V character, which is indicative of an Ohmic-like interface. This is in agreement with transport results as well as the theory for constructing band alignments in oxide systems

These findings provide experimental evidence that support the theory for band alignment of multicomponent oxides, which predicts that the SrTaO$_3$/SrTiO$_3$ system forms an Ohmic junction, whereas SrVO$_3$/SrTiO$_3$ forms a Schottky junction. This observation is critical to expanding into all oxide-electronics which can make use of the great diversity of phenomena known in the multicomponent oxides. Specifically, this provides a route to design energy level alignments that can be used to dictate charge transfer and local electric fields. This ability establishes a clear path to new levels of functional tunability in oxide electronics. Key examples for future work center on materials with traditional functionality such as semiconductors and more exotic materials. For semiconductor devices, there is a particular importance of $SrTiO_3$,[4] which has a large and highly tunable dielectric constant[41] and lattice matching to Si and other semiconductors,[3] but also other dielectrics such as $(Sr,Ba)TiO_3$[42] and $Sr_{n+1}Ti_nO_4$,[43] as well as $SrZrO_3$ and $SrHfO_3$ which also have highly-tunable dielectric responses. Here, for $SrTiO_3$, $Sr(V-Ta)O_3$ are likely good choices to engineer desired interfacial properties, where other perovskite metals ($SrRuO_3$ and $SrIrO_3$) are predicted to only form Schottky junctions. Interestingly, theory predicts that for $SrZrO_3$ only $SrTaO_3$ is Ohmic, while for $SrHfO_3$, $Sr(V-Ta)O_3$ are all Schottky-like. Extending this scheme outside of transition metals to include the Sn-based perovskites such as $BaSnO_3$ are of significant interest, since the extended $p$-bands of Sn give rise to relatively wide bandgaps and large room-temperature mobility, which is critical for room temperature applications, especially in the realm of power electronics.[44,45] Creating well-controlled high-quality interfaces and epitaxial contacts to the stannates is an open question for future work.

Beyond these traditional functionalities a host of other useful properties are known, for which creating and controlling the interfacial charge transfer is crucial. Novel magnetic states abound across the transition metals, which can be highly tunable via interfacial charge transfer.[18] A key example being $SrRuO_3$ ($\varepsilon_p$=-3.40 eV), which is a ferromagnetic metal and would be an ideal candidate for spin-injection into a high mobility 2DEG in $SrTiO_3$, $SrZrO_3$ and $SrHfO_3$ ($\varepsilon_p$=-4.00,-6.23,-7.43 eV, respectively). However, predictions show that these interfaces are Schottky-like. Therefore, additional work to modify the interface is required to realize such a device. While much work has been done to engineer chiral magnetic interactions at oxide interfaces, it is predominately focused on hybridizing the strong spin-orbit coupling in the metal $SrIrO_3$ with magnetic insulators such as $SrMnO_3$ and $LaMnO_3$, where the electronic states at the interfaces are required to break inversion symmetry[46]. Magnetic interactions may be easily controlled by engineering charge transfer across the interface to drive symmetry breaking in $SrIrO_3$. Lastly, many perovskites are ideal platforms for energy-efficient neuromorphic computing that emerge from a strong interplay of metal-insulator transitions and externally controlled oxygen migration, as demonstrated in the rare-earth nickelates.[47] Such processes rely on overcoming built in energy barriers at interfaces and therefore represent a wide array of possible tunability through designing interfacial charge transfer that may be enabled using informed oxide junction selection.

To conclude, a necessary step towards oxide electronics is the ability to design, tailor, and create charge transfer at epitaxial oxide/oxide interfaces. Towards this goal, we show that Ohmic- and Schottky-like interfaces can be designed and realized via a novel band alignment scheme developed for the multifunctional transition metal oxides. This is demonstrated for the important class of perovskite metals ($Sr(V-Ta)O_3$) interfaced with the multifaceted semiconducting oxide, $SrTiO_3$. Here, the charge associated with occupied states in $SrTiO_3$ is transferred to the $SrVO_3$ while $SrTaO_3$ creates an accumulation of charge within the $SrTiO_3$ interfacial layers. As only a handful of perovskites are good metals, the $SrVO_3$-$SrTaO_3$ series are a critical class for providing top or bottom gates, as well as highly-conductive contact materials. Also, due to the strong electron-electron correlations these materials are highly transparent which make them relatively straight-forward building blocks for oxide-electronics. Going forward we have discussed key examples and open questions where understanding band alignment and charge transfer at complex oxide interfaces is critical to designing and utilizing their diverse functionalities. Mastering these is key towards realizing and overcoming several challenges facing the continued evolution of electronics.

## Acknowledgements

This work was supported by the U.S. Department of Energy (DOE), Office of Science, Basic Energy Sciences (BES), Materials Sciences and Engineering Division (growth, electronic characterization, and first principles calculations), the U.S. DOE, Office of Science, National Quantum Information Science Research Centers, Quantum Science Center (structural characterization).

## Conflict of Interest

The author declares no competing financial interest.

## Supporting Information

See Supporting Information for experiment and theorical methods as well as discussion of multicarrier fit. Supporting information includes Refs. [26,28,35,48,49].

**Figures**

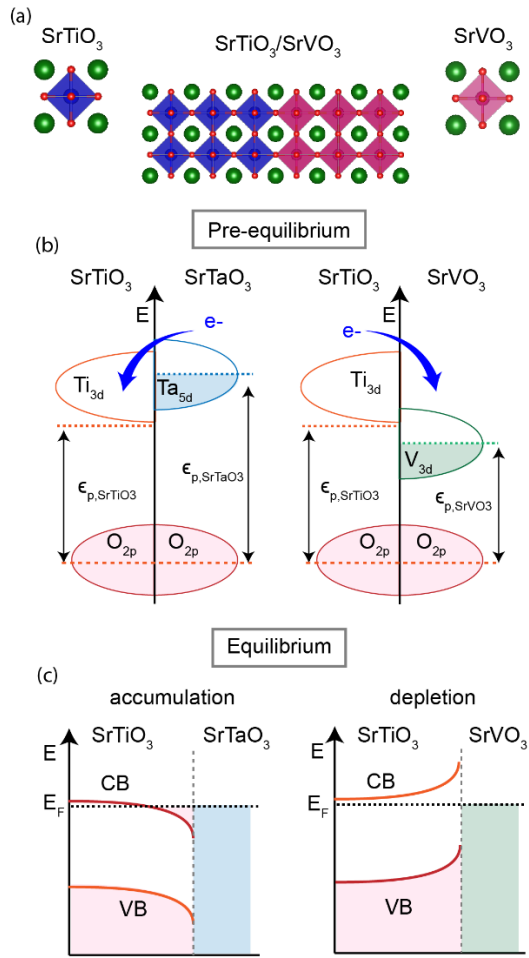

**Figure 1** (a) Crystal structures of perovskite compounds SrVO$_3$ and SrTiO$_3$, and the corresponding heterojunction. (b) Band alignment in Sr$B$O$_3$-SrTiO$_3$ heterojunction showing different charge transfer directions of SrVO$_3$ vs SrTaO$_3$ when interface with SrTiO$_3$ based on their bulk energy states and the continuum oxygen state boundary condition as predicted in Ref. [19]. (c) Schematic of the corresponding band bending that occurs in Sr$B$O$_3$-SrTiO$_3$ heterojunctions after equilibrium is established, which shows the formation of an Ohmic-type interface in SrTaO$_3$/SrTiO$_3$ and a Schottky-type interface in SrVO$_3$/SrTiO$_3$.

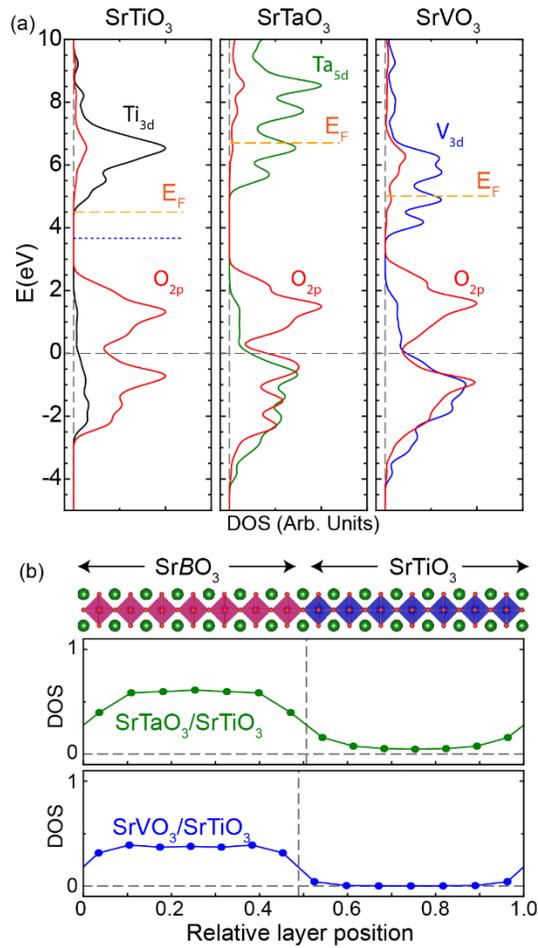

**Figure 2** (a) Energy dependent density of states (DOS) for bulk $SrTiO_3$ (left), $SrTaO_3$ (center) and $SrVO_3$ (right), where the energy has been shifted to align the O2$p$ states at $E = 0$ eV which is denoted by a dashed gray line. The red curves are the O2$p$ states, and the black/green/blue curves are the transition metal $d$-states, as indicated. The orange line indicates the Fermi energy (the dotted blue line for $SrTiO_3$ indicates the Fermi energy in the intrinsic limit). (b) Spatially resolved layer-by-layer # electrons for superlattice systems. Data points indicate B-site positions within a supercell as indicated by the crystal structure model vertically above. Left of the dashed vertical line is the $SrTaO_3$ (top panel)/$SrVO_3$ (bottom panel) side of the interface, and the right side is the $SrTiO_3$. The non-zero number of electrons on the $SrTiO_3$ indicates the presents of a 2DEG due to charge transfer into the $SrTiO_3$ for the case of $SrTaO_3$, but not for $SrVO_3$.

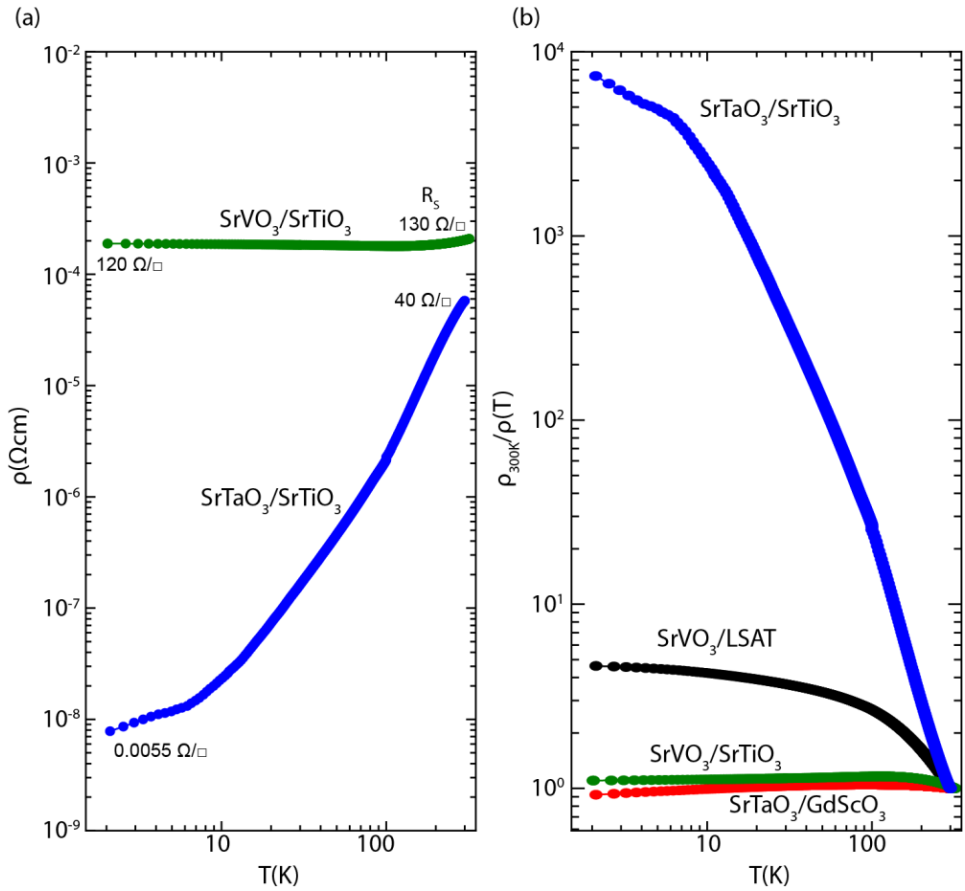

**Figure 3** (a) Temperature dependence of longitudinal resistivities for $SrVO_3/SrTiO_3$ (16 nm) and $SrTaO_3/SrTiO_3$ (14 nm), and indicated are the sheet resistance, $R_S$, values at room temperature and 2 K (see Supporting Information Figure S4 for a plot of $R_S$ vs $T$). (b) $\rho_{300K}/\rho(T)$ for $SrVO_3/SrTiO_3$ and $SrTaO_3/SrTiO_3$ shown in (a) as well as films grown on highly insulating $SrVO_3/LSAT$ and $SrTaO_3/GdScO_3$ demonstrating only $SrTaO_3$-$SrTiO_3$ exhibits a high residual resistivity ratio of order of 10,000 due to the Ohmic type interface.

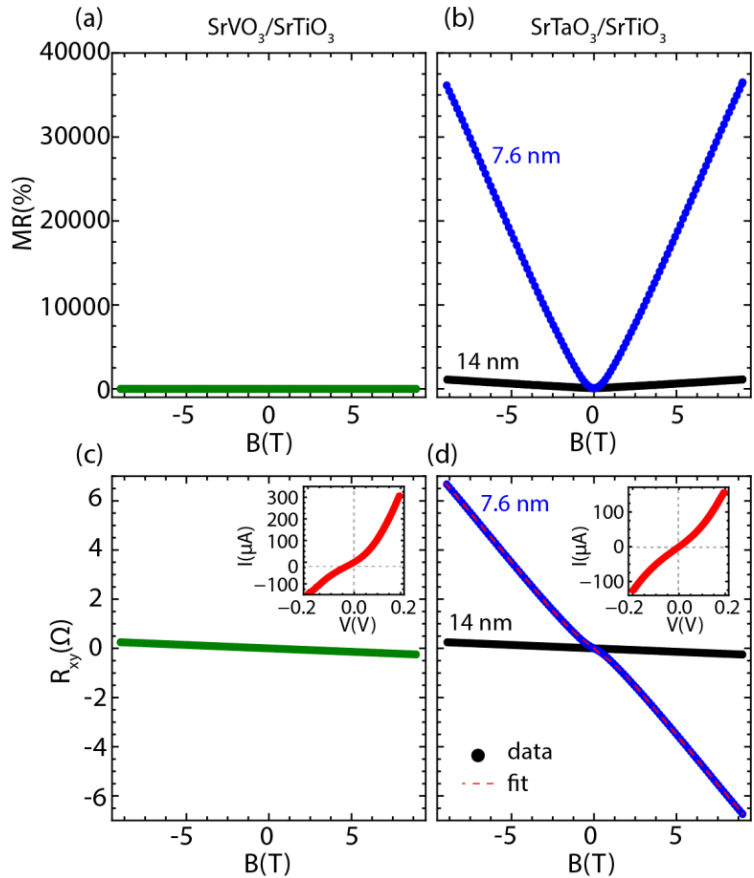

**Figure 4** Transport data of selected SrVO$_3$/SrTiO$_3$ (left panels) and SrTaO$_3$/SrTiO$_3$ (right panels). (a,b) MR showing MR ~ 0.1 % in SrVO$_3$/SrTiO$_3$ (a) but nearly 40,000 % in SrTaO$_3$/SrTiO$_3$ (b). (c,d) Hall effect, where SrVO$_3$ shows a linear dependence (c) whereas SrTaO$_3$/SrTiO$_3$ shows non-linear behavior (d). See Supporting Information for multicarrier fitting and detailed analysis and plots of *dR$_{xy}$/dB* which highlight the non-linear Hall effect. Insets in (c,d) are current voltage (I-V) curves for similar samples grown on Nb-doped SrTiO$_3$ with doping level 0.5%.

# Supporting Information

# Design and realization of Ohmic and Schottky interfaces for oxide electronics

Jie Zhang[1], Yun-Yi Pai[1], Jason Lapano[1], Alessandro R. Mazza[1], Ho Nyung Lee[1], Rob Moore[1], Benjamin J. Lawrie[1], T. Zac Ward[1], Gyula Eres[1], Valentino R. Cooper[1#], Matthew Brahlek[1]*

[1]Oak Ridge National Laboratory, Oak Ridge, Tennessee, U.S.A.

Email: *brahlekm@ornl.gov, #coopervr@ornl.gov

**Film growth**

Epitaxial $SrVO_3$ and $SrTaO_3$ thin films were grown on $SrTiO_3$ (Crystec Corporation) using pulsed laser deposition (PLD). Prior to growth the substrates were chemically etched with buffered hydrofluoric acid before annealing in air at 1000 ºC for 2 hours, to achieve nominal $TiO_2$ termination and atomically flat surfaces. A KrF excimer laser ($\lambda$ = 248 nm) was used to ablate stoichiometric targets at a repetition rate of 5 Hz and fluence of 0.5 J/cm$^2$, an oxygen partial pressure of $4 \times 10^{-6}$ Torr, and a growth temperature of 650 ºC. Crystallinity was examined with X-ray diffraction (XRD) using a four-circle diffractometer (Panalytical Corporation) with Cu-k$_{\alpha 1}$ radiation as shown in Figure S1. Reciprocal space maps demonstrate that the films were fully strained. The out-of-plane lattice constants were extracted from the 002 peak position ($c$ = 3.80 Å for $SrVO_3$ and $c$ = 4.01 Å for $SrTaO_3$). Film thickness was confirmed by X-ray reflectivity (XRR). Transport measurements were performed in a physical property measurement system (PPMS, Quantum Design) using van der Pauw sample configuration by soldering indium contacts.

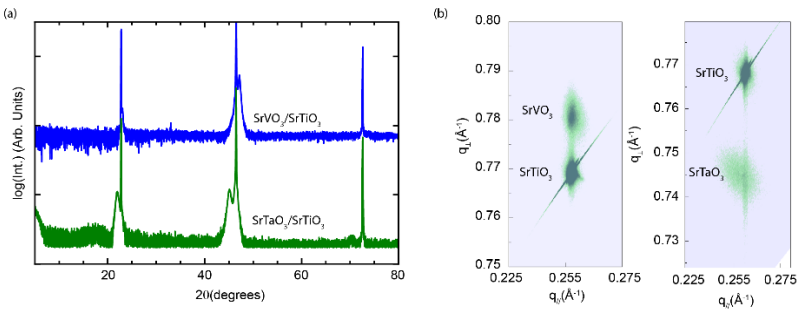

**Figure S1**. X-ray diffraction analysis of $SrVO_3/SrTiO_3$ and $SrTaO_3/SrTiO_3$ showing $2\theta$-$\theta$ scans about the 00$L$ reflections (a), as well as reciprocal space maps about the 103 set of reflections (b).

**First Principles Calculations**

All density functional theory calculations were performed using the QUANTUM ESPRESSO simulation code (v 6.4.1) [1] with the local density approximation (LDA) for exchange and correlation and ultrasoft pseudopotentials. The Sr 4s4p5s, Ti 3p3d4s, V 4s3d, Nb 4p4d5s, Ta 5d6s and O 2s2p electrons were treated as valence electrons. A 500 eV cutoff and an 8×8×8 or 8×8×1 Monkhorst-Pack $k$-point mesh was employed for the bulk or superlattice calculations, respectively. A Hubbard U of 5.0 eV for Ti, V, Nb, and Ta d-states was found to be appropriate for all calculations. Similar U values were found to give a reasonable description of the electronic and structural properties of related 2DEG model systems. [2–4] The computed bulk $SrVO_3$, $SrNbO_3$, $SrTaO_3$, and $SrTiO_3$ cubic lattice constants of 3.757, 3.987, 3.971, and 3.879 Å, respectively, are in typical LDA agreement with the experimental values of 3.843, 4.024, 4.00 (the bulk lattice parameter of $SrTaO_3$ is not well known), and 3.900 Å, respectively. For all interface calculations, a 1×1×14 perovskite unit cell superlattice was used. With a stoichiometry equivalent to 7 layers of $SrTiO_3$ and 7 layers of $SrBO_3$ (where $B$=V, Nb, and Ta). The in-plane lattice constants $a$ and $b$ were constrained to the theoretical value of $SrTiO_3$ (3.879 Å), while the out-of-plane lattice vector $c$ was optimized within the P4mm space group. Ionic coordinates were optimized until all Hellman-Feynman forces were less than 5 meV/Å. Reported conduction electrons were computed from states from $E_F$ to the gap defined by the nominally O2$p$ states.

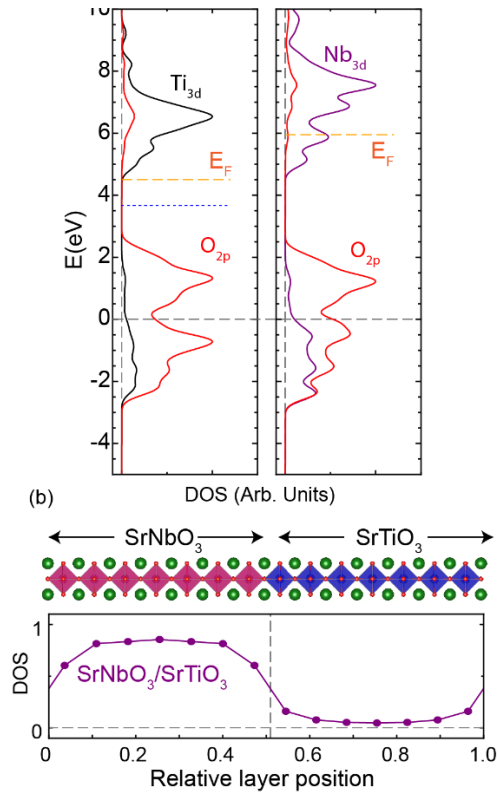

**Figure S2**. (a) Energy dependent density of states (DOS) for bulk $SrTiO_3$ (left), $SrNbO_3$ (right) where the energy has been shifted to align the O2$p$ states at $E = 0$ eV which is denoted by a dashed gray line. The red curves are the O2$p$ states, and the black/purple curves are the transition metal $d$-states, as indicated. The orange line indicates the Fermi energy (the dotted blue line for $SrTiO_3$ indicates the Fermi energy in the intrinsic limit). (b) Spatially resolved layer-by-layer # electrons for superlattice systems. Data points indicate Nb-site positions within a supercell as indicated by the crystal structure model vertically above. Left of the dashed vertical line is the $SrNbO_3$ side of the interface, and the right side is the $SrTiO_3$. The non-zero number of electrons on the $SrTiO_3$ indicates the presents of a 2DEG due to charge transfer into the $SrTiO_3$ from $SrNbO_3$.

**Multicarrier transport model**

A multicarrier model for the Hall effect can be used to gain insight into the origin of the observed non-linearity in the SrTaO$_3$/SrTiO$_3$ system. The mathematical functions describing the Hall resistance and magnetoresistance of a multiband metal as a function of the areal carrier densities ($n_i$) and mobilities ($\mu_i$) can be derived by considering parallel channels. As such, the conductance is simply the summation of the individual components:

$$G_{xx} = \sum_i \frac{e|n_i|\mu_i}{1+\mu_i^2 B^2}$$

$$G_{xy} = \sum_i \frac{en_i\mu_i^2 B}{1+\mu_i^2 B^2}$$

where $e$ is the electron charge and $B$ is the magnetic field. Then the sheet and the Hall resistivities are given, respectively, by

$$R_{xx} = \frac{G_{xx}}{G_{xx}^2 + G_{xy}^2},$$

and

$$R_{xy} = \frac{G_{xy}}{G_{xx}^2 + G_{xy}^2}.$$

The conductivities and resistivities are related to the conductance and resistances as $\sigma = G/thickness$ and $\rho = R*thickness$.

This provides a set of equations that can be used to fit experimental Hall effect data, with 2 parameters ($n$ and $\mu$) per conductance channel. For the case of SrTaO$_3$ on SrTiO$_3$ the resistivity versus temperature analysis in the main text shows that the SrTiO$_3$ substrate contributes to the transport with a carrier density in the range of $10^{16}$-$10^{17}$ cm$^{-3}$, which implies that there are at least two conductance channels. A third channel emerges due to populating the interfacial SrTiO$_3$, which is consistent with the band alignment prediction as well as experimentally consistent with the nonlinearity observed in the Hall effect. This 3-channel model requires 6 independent parameters. However, the well-established transport phenomena of SrTiO$_3$, and the independent measurements of SrTaO$_3$ on highly-insulating GdScO$_3$ enables us to confine to 3 parameters, as follows. Based on the analysis of resistivity versus temperature in the main text shows that SrTiO$_3$ in the estimated carrier density range has a mobility of the order of 1,000 cm$^2$V$^{-1}$s$^{-1}$, which we take to be roughly this value. SrTaO$_3$ has a carrier density roughly $1.7 \times 10^{22}$ cm$^{-3}$ (1 electron per unit cell), and a mobility of order 1 cm$^2$V$^{-1}$s$^{-1}$. As such, we let the carrier density of the SrTiO$_3$ be a fit parameter as well as the carrier density and mobility of the 2DEG state that is responsible for the hole-like behavior. This reduces 6 parameters to 3 parameters and enables us to better understand the unusual behavior observed for the SrTaO$_3$/SrTiO$_3$ system. We stress that due to the complex band structure and scattering processes the multicarrier analysis only gives a semi-qualitative understanding of the different transport channels, and the estimated error in the stated numbers is likely large. As such, we fit the experimental data only when there is a strong non-linearity (<30 K for the 7.6 nm SrTaO$_3$/SrTiO$_3$ film) and estimate how the various transport parameters change which describes the higher temperature regime and thicker films.

First, as discussed in the main text and shown in Figure 4, a strong non-linear Hall effect is found for thin samples at low temperature. For thicker samples and higher temperatures, the non-linearity is reduced. For the 7.6 nm sample the non-linearity is absent at 300 K, and with reducing temperature is peaked around 8 K, and with further reducing temperature the magnitude steadily drops down to 2 K. As such, we first fit the 2 K data, with the results shown in Figure S3 and the parameters in Table S1. The fit agrees excellently with the experimental data, which yields an SrTiO$_3$ carrier density in the range of $10^{16}$ cm$^{-3}$ and an areal carrier density of the 2DEG state of order of $10^{13}$ cm$^{-2}$ (to extract the volume carrier density a thickness of 5 nm was assumed) and a 2DEG mobility of order 3700 cm$^2$V$^{-1}$s$^{-1}$. These values are in line with typical values measured for SrTiO$_3$ as well as 2DEGs, and further represents a reasonable amount of charge that may be transferred across an interface.

Secondly, to understand why the overall slope drops for large film thicknesses we note the following. First, the areal carrier density of the SrTaO$_3$ increases, but also the carrier density of the SrTiO$_3$ substrate increases as well. This is due to more oxygen vacancies being produced due to the longer dwell time at the growth temperature required. Hence, in Figure S3(a), we simulated this by increasing both the areal density of SrTaO$_3$ as well as SrTiO$_3$. What can be seen is the overall Hall slope drops with increasing SrTiO$_3$ carrier density. This exactly reproduces the experimental trend.

Thirdly, at higher temperatures the slope of $R_{xy}$ for the 7.6 nm film drops. This is most likely driven by a change to the mobilities in bulk SrTiO$_3$ and 2DEG systems, where the mobilities are highly temperature dependent. As such, in Figure S3(b) we simply lowered the mobility of both the SrTiO$_3$ and the 2DEG state by a factor of 7, where the mobilities are of order of 100. At this value the simulations shows that this reduces the slope of $R_{xy}$ and eliminates the non-linearity, which exactly agrees with the room-temperature experimental data shown in this figure. We note that these mobilities are higher than typical values at room temperature (typically of order of 10 cm$^2$V$^{-1}$s$^{-1}$), and, in fact, as noted below, this may be confounded with a change in carrier density of SrTiO$_3$ at elevated temperatures since the Fermi energy is smaller than the temperature scale). This again highlights the semi-qualitative nature of our model and stresses that these changes should only be interpreted as order-of-magnitude estimations.

Finally, the non-monotonic change of the Hall slope with reducing temperature for the 7.6 nm SrTaO$_3$/SrTiO$_3$ is very unusual. To see this, compare the data shown in Figure S3(a) at 2 K to the data in Figure S3(c) which was measured at 8 K. To estimate the change in parameters, we performed the same fitting procedure, which, as shown in Table S1, shows an increase in the carrier density of the 2DEG ($1.5\times10^{13}$ cm$^{-2}$ → $2.2\times10^{13}$ cm$^{-2}$), a slight reduction in the 2DEG mobility (3700 cm$^2$V$^{-1}$s$^{-1}$ → 3300 cm$^2$V$^{-1}$s$^{-1}$), and the SrTiO$_3$ carrier density is nearly constant. A second fit is shown where the 2DEG values were fixed at those of the 2 K fit, and only the SrTiO$_3$ parameters were varied. The rational is that at a very dilute density these parameters are known to both be dependent on temperature (see for example Figure 3 in Ref. [5]), which can be rationalized since the Fermi energy is of order the temperature (i.e. $E_F = \hbar^2(3\pi^2 n)^{2/3}/(2m^*) \approx 2$ K, where $\hbar$ is Planck's reduced constant, and $m^*$ is the effective mass, which is well known to be of order the electron mass for SrTiO$_3$). Here the fit is equally good and shows that the carrier density and mobility slightly decrease in going from 2 K to 8 K, which is consistent with the magnitude of changes observed in bulk SrTiO$_3$. However, the order-of-magnitude of the fit parameters are likely correct, yet the finer changes are likely within the error bars (since two models with different parameters yield excellent fits); therefore, it is likely not meaningful to scrutinize the physical origins of the relative changes.

Overall, the multicarrier analysis shows that for the SrTaO$_3$/SrTiO$_3$ that the interface must be populated by charge transfer from the SrTaO$_3$. More specifically, the transport can only be explained by the high carrier density SrTaO$_3$ and two high mobility states, which can only be the lightly doped SrTiO$_3$ substrate and an interfacial 2DEG. In contrast, the SrVO$_3$/SrTiO$_3$ system clearly shows no charge transfer, and the transport is solely through the SrVO$_3$. This together is fully supported by the band alignment analysis, which predicts that SrTaO$_3$ forms an Ohmic-type interface with SrTiO$_3$ and SrVO$_3$ forms a Schottky-type interface.

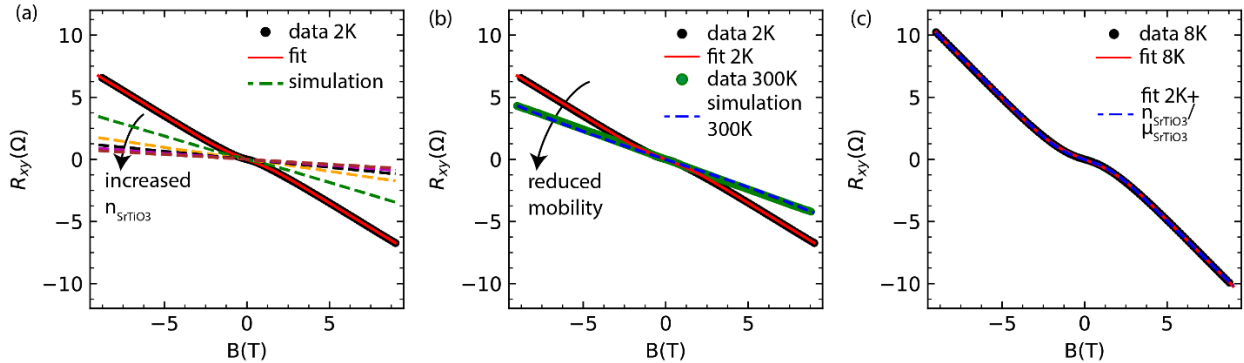

**Figure S3.** Data, multicarrier fits, and simulations for the 7.6 nm SrTaO$_3$ for various temperatures. (a) Data taken at 2 K, fit, and simulations with increasing the SrTiO$_3$ carrier density which reproduces the trend for thicker SrTaO$_3$/SrTiO$_3$ as shown in Figure 4 of the main text. (b) Data taken at 2 K and fit, as well as data taken at 300 K. Simulation corresponding to the parameters of the 2 K fit with mobility reduced, which reproduces the 300 K data. (c) Data taken at 8 K, which shows a larger $R_{xy}$ relative to 2 K, which is fit as in (a), red, as well as by taking the 2DEG fit parameters from 2 K then varying the SrTiO$_3$ carrier density and mobility.

|  |  | SrTiO$_3$ |  | SrTaO$_3$ |  | 2DEG |  |  |
|---|---|---|---|---|---|---|---|---|
| Panel/color | Type | n(cm$^{-3}$) | μ (cm$^2$V$^{-1}$s$^{-1}$) | n(cm$^{-3}$) | μ (cm$^2$V$^{-1}$s$^{-1}$) | n(cm$^{-3}$) | n(cm$^{-2}$) | μ (cm$^{-2}$V$^{-1}$s$^{-1}$) |
| (a)/red | fit | -8.0E+15 | 1000 | -1.7E+22 | 1 | 3.0E+19 | 1.5E+13 | 3700 |
| (a)/green | simulation | -1.6E+16 | 1000 | -1.7E+22 | 1 | 3.0E+19 | 1.5E+13 | 3700 |
| (a)/orange | simulation | -3.2E+16 | 1000 | -1.7E+22 | 1 | 3.0E+19 | 1.5E+13 | 3700 |
| (a)/black | simulation | -4.8E+16 | 1000 | -1.7E+22 | 1 | 3.0E+19 | 1.5E+13 | 3700 |
| (a)/magenta | simulation | -6.4E+16 | 1000 | -1.7E+22 | 1 | 3.0E+19 | 1.5E+13 | 3700 |
| (a)/brown | simulation | -8.0E+16 | 1000 | -1.7E+22 | 1 | 3.0E+19 | 1.5E+13 | 3700 |
| (b)/red | fit | -8.0E+15 | 1000 | -1.7E+22 | 1 | 3.0E+19 | 1.5E+13 | 3700 |
| (b)/blue | simulation | -1.0E+16 | 140 | -1.7E+22 | 1 | 4.1E+20 | 2.0E+14 | 500 |
| (c)/red | fit | -5.0E+15 | 1000 | -1.7E+22 | 1 | 4.3E+19 | 2.2E+13 | 3300 |
| (c)/blue | fit | -5.0E+15 | 860 | -1.7E+22 | 1 | 3.0E+19 | 1.5E+13 | 3700 |

**Table S1.** Parameters of the multicarrier fit and simulations shown in Figure S3. The carrier density for the channel associated with a 2DEG is shown in 3D density (assuming ~5 nm thickness).

**Additional transport data**

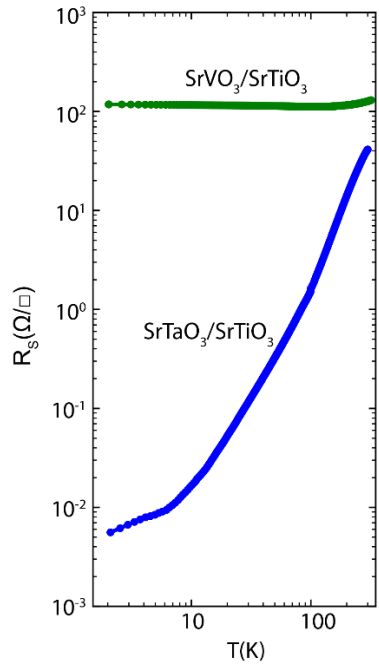

**Figure S4.** Temperature dependence of longitudinal sheet resistance for $SrVO_3/SrTiO_3$ (16 nm) and $SrTaO_3/SrTiO_3$ (14 nm).

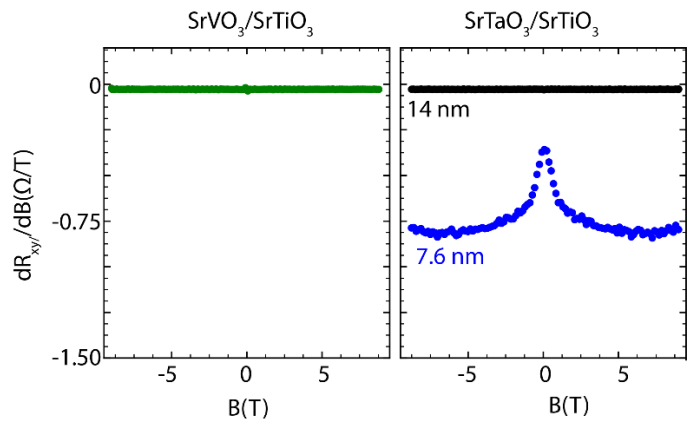

**Figure S5.** Derivative for the Hall effect ($dR_{xy}/dB$), where $SrVO_3$ is constant (c) whereas $SrTaO_3/SrTiO_3$ shows non-linear behavior. This data corresponds to the same samples as show in Figure 4 of the main text.